\RequirePackage[2020-02-02]{latexrelease}
\documentclass[aps,prd,groupedaddress,graphicx,nofootinbib]{revtex4}
\usepackage{amsmath,amssymb,graphics,graphicx,color,epsf}
\usepackage{subfigure}
\usepackage[scanall]{psfrag}  
\usepackage{tikz}

\begin{document}

\title{Uniform rate inflation on the brane}

\author{Chia-Min Lin$^{1}$}
\author{Rei Tamura$^{2}$}
\author{Keiko I. Nagao$^{2}$}

\affiliation{$^{1}$Fundamental General Education Center, National Chin-Yi University of Technology, Taichung 411030, Taiwan, R.O.C.}
\affiliation{$^{2}$Department of Physics, Faculty of Science, Okayama University of Science, 1-1 Ridaicho, Okayama, 700-0005, Japan}



\begin{abstract}
We propose a model of uniform rate inflation on the brane. The potential is given by a hyperbolic cosine function plus a negative cosmological constant. The equation of motion is solved analytically without using slow-roll approximation. The result is that the inflaton field is rolling at a constant speed. The prediction for cosmological perturbations depends on the field value at the end of inflation. The experimental constraints could be satisfied in the parameter space.

\end{abstract}
\maketitle
\large
\baselineskip 18pt
\section{Introduction}
\label{sec1}

Cosmic inflation \cite{Starobinsky:1980te, Guth:1980zm, Linde:1981mu} is arguably the standard scenario of the very early universe. The single idea solves almost all the problems of the traditional hot big bang model, such as the flatness problem, the horizon problem, and the unwanted relics problem. The quantum fluctuations during inflation are amplified to cosmological scales and could generate primordial tensor and scalar perturbations constrained by experimental observations such as cosmic microwave background (CMB) observed by Planck \cite{Planck:2018jri}.

However, many inflation models exist, and there is no consensus about the best model. Occam's razor recommends simplicity.
The simplest inflation models are described by a scalar field $\phi$ called the inflaton field with a potential $V(\phi)$. 
Presumably, the energy density of the universe $\rho$ during inflation is dominated by the energy density of $\phi$ given by
\begin{equation}
\rho=\frac{1}{2}\dot{\phi}^2+V(\phi).
\label{rho}
\end{equation}
Conventionally, the Friedmann equation of general relativity in a flat, homogeneous, and isotropic universe is 
\begin{equation}
3H^2=\rho,
\end{equation}
where $H \equiv \dot{a}/a$ is the Hubble parameter with $a$ the scale factor. Here and throughout this work, we set the reduced Planck mass to unity, that is, $M_P \equiv 1/\sqrt{8\pi G}=1$.
The equation of motion of the inflaton field is 
\begin{equation}
\ddot{\phi}+3H\dot{\phi}+V^\prime(\phi)=0.
\label{eom}
\end{equation}
In \cite{Lin:2023xgs}, one of the authors proposed a model called uniform rate inflation where the inflaton field has the potential
\begin{equation}
V(\phi)=\frac{3\lambda^2\phi^2}{4}-\frac{\lambda^2}{2}.
\label{unipo}
\end{equation}
The potential consists of a quadratic term and a related negative cosmological constant.
In this model, the equation of motion is solved analytically as
\begin{equation}
\dot{\phi}=-\lambda.
\label{uni}
\end{equation}
The inflaton field is rolling at a uniform rate; thus, it is called uniform rate inflation.
Note that there is no need to use slow-roll approximation to analyze the equation of motion. 
The solution was found through a study of quantum cosmology in the framework of de Broglie-Bohm theory \cite{Lin:2023sza}.
The time evolution can be easily integrated to
\begin{equation}
\phi=-\lambda t+d,
\label{phi}
\end{equation}
where $d$ is the integration constant. 
This model is simple and neat. This work aims to extend uniform rate inflation to the braneworld and discover the predictions.

\section{braneworld cosmology}

The idea of the braneworld proposes that our four-dimensional world is a four-dimensional hypersurface (3-brane) embedded in a higher dimensional bulk \cite{Arkani-Hamed:1998sfv}. In this scenario, ordinary matter is confined to our world, while gravity is in the whole $(4+n)$-dimensional spacetime. Such models are motivated by strongly coupled string theories \cite{Horava:1996ma, Antoniadis:1998ig}. To have sufficiently small extra dimensions not excluded by short-distance gravitational measurements, we should have $n \geq 2$ from the argument of dimensional reduction. However, Randall and Sundrum discovered that the $n=1$ case is still allowed because of the gravitational effects of a brane on the extra dimension \cite{Randall:1999ee, Randall:1999vf}. 
If one concerns cosmological evolution of the universe, the Friedmann equation is modified to \cite{Cline:1999ts, Csaki:1999jh, Binetruy:1999ut, Binetruy:1999hy, Freese:2002sq, Freese:2002gv, Maartens:1999hf}
\begin{equation}
3H^2=\rho \left( 1+\frac{\rho}{2\Lambda} \right),
\label{brane}
\end{equation}
where the brane tension $\Lambda$ is related to the 
5-dimensional Planck mass $M_5$ as

\begin{equation}
\Lambda=\frac{3M_5^6}{32\pi^2}.
\end{equation}
We recover the conventional Friedmann equation if $\rho \ll 2\Lambda$ (like our current universe).
The nucleosynthesis limit implies that $\Lambda \gtrsim (1\mbox{ MeV})^4\sim (4.1\times 10^{-22})^4$. A more stringent constraint can be obtained by requiring the theory to reduce to Newtonian gravity on scales larger than $1$ mm; this gives $M_5 \gtrsim 10^5\mbox{ TeV}$. which corresponds to $\Lambda \gtrsim 4.2 \times 10^{-64}$ \cite{Maartens:1999hf}.
We consider a braneworld described by Eq.~(\ref{brane}) in the following discussion.

\section{uniform rate inflation on the brane}

We seek a potential $V(\phi)$ in the braneworld scenario, which has the solution of the equation of motion given by Eq.~(\ref{uni}).
Note that $\ddot{\phi}$=0, therefore the equation of motion from Eq.~(\ref{eom}) is
\begin{equation}
-3\lambda H+V^\prime(\phi)=0,
\label{emo}
\end{equation}
where we have used Eq.~(\ref{uni}).
By using Eqs.~(\ref{rho}), (\ref{uni}), (\ref{brane}), and (\ref{emo}), we can write
\begin{equation}
(V(\phi)^\prime)^2=3\lambda^2 \left[ \frac{\lambda^2}{2}+V(\phi)+\frac{\left(\frac{\lambda^2}{2}+V(\phi)\right)^2}{2\Lambda} \right].
\end{equation}
Interestingly, this equation can be integrated to obtain
\begin{equation}
V=\frac{\Lambda c}{2}e^{\sqrt{\frac{3\lambda^2}{2\Lambda}}\phi}+\frac{\Lambda}{2c}e^{-\sqrt{\frac{3\lambda^2}{2\Lambda}}\phi}-\Lambda -\frac{\lambda^2}{2},
\end{equation}
where $c$ is an arbitrary (suitably rescaled) integration constant. Different choices of the constant $c$ correspond to the field redefinition, which we are free to make. Therefore, without loss of generality, we choose $c=1$ to obtain 
\begin{equation}
V(\phi)=\Lambda \cosh\sqrt{\frac{3\lambda^2}{2\Lambda}}\phi -\Lambda -\frac{\lambda^2}{2}.
\label{pob}
\end{equation}
This would be the model considered in the following discussion.
In retrospect, it is pretty remarkable that such simple potential admitted an analytical solution given by Eq.~(\ref{uni}) for the complicated equation of motion provided by Eqs.~(\ref{rho}), (\ref{eom}), and (\ref{brane}). Note that we do not need slow-roll approximation here. The stability of the solution is discussed in the Appendix.

It is interesting to note that when $\sqrt{\frac{3\lambda^2}{2\Lambda}}\phi \ll 1$, we can expand the potential of Eq.~(\ref{pob}) by using 
\begin{equation}
\cosh \theta= 1+ \frac{\theta^2}{2}+\cdots.
\end{equation} 
The result is that we recover Eq.~(\ref{unipo}) with the predictions of primordial perturbations given in \cite{Lin:2023xgs}. Therefore, Eq.~(\ref{pob}) may be regarded as the generalization of the uniform rate inflation given by Eq.~(\ref{unipo}) to the braneworld.
From Eqs.~(\ref{rho}) and (\ref{pob}), we have
\begin{equation}
\rho=\Lambda \left( \cosh \sqrt{\frac{3\lambda^2}{2\Lambda}}\phi-1 \right) \simeq \frac{\Lambda}{2} \left( \sqrt{\frac{3\lambda^2}{2\Lambda}}\phi \right)^2.
\end{equation}
Therefore, from Eq.~(\ref{brane}), we can regard this as the low-energy limit where the brane high-energy corrections become unimportant.

Since the potential is an even function of $\phi$, we can assume $\phi>0$ in the following discussions without loss of generality. The Hubble parameter can be obtained from Eq.~(\ref{emo}) as
\begin{equation}
H=\frac{\dot{a}}{a}=\sqrt{\frac{\Lambda}{6}}\sinh \sqrt{\frac{3\lambda^2}{2\Lambda}}\phi.
\label{a}
\end{equation}
This can be integrated to obtain the scale factor either as a function of time $t$ or as a function of $\phi$, which is related through $t$ via Eq.~(\ref{phi}) as
\begin{equation}
a=e^{-\frac{\Lambda}{3\lambda^2}\cosh \sqrt{\frac{3\lambda^2}{2\Lambda}}\phi} \equiv e^\alpha,
\label{alp}
\end{equation}
(up to an irrelevant integration constant) where we have defined a parameter $\alpha$.
In slow-roll inflation, the scale factor during inflation is described as quasi-exponential or quasi-de Sitter expansion. Our model has a closed form for the scale factor, which shows precisely how the scale factor evolves.

The field value $\phi=\phi_e$ at the end of inflation is obtained by solving $\ddot{a}=0$, which gives
\begin{equation}
\cosh \sqrt{\frac{3\lambda^2}{2\Lambda}}\phi_e=\frac{3\lambda^2}{2\Lambda}+\sqrt{\frac{9\lambda^4}{4\Lambda^2}+1}.
\end{equation}
When $\phi>\phi_e$, inflation is happening with $\ddot{a}>0$. 
As discussed in \cite{Lin:2023xgs}, after inflation, the inflaton field does not oscillate but keeps on going with a uniform rate if the potential remains the same, and the universe tends to collapse shortly.
We assume that somehow reheating happens after $\phi=\phi_e$ and recovers hot big bang. 
The inflaton field value $\phi_{60}$ at the number of e-folds $N=60$ (which corresponds to horizon exit or CMB scale) can be obtained by solving $\Delta \alpha=60$ that is
\begin{equation}
-\frac{\Lambda}{3\lambda^2}\left( \frac{3\lambda^2}{2\Lambda}+\sqrt{\frac{9\lambda^4}{4\Lambda^2}+1} \right)+\frac{\Lambda}{3\lambda^2}\cosh \sqrt{\frac{3\lambda^2}{2\Lambda}}\phi_{60}=60.
\label{he}
\end{equation}
This gives
\begin{equation}
\cosh \sqrt{\frac{3\lambda^2}{2\Lambda}}\phi_{60}=\frac{363\lambda^2}{2\Lambda}+\sqrt{\frac{9\lambda^4}{4\Lambda^2}+1}.
\label{co}
\end{equation}
A useful identity is
\begin{equation}
\cosh^2 \sqrt{\frac{3\lambda^2}{2\Lambda}}\phi_{60}-1=\frac{65889 \lambda^4}{2\Lambda^2}+\frac{363\lambda^2}{\Lambda}\sqrt{\frac{9\lambda^4}{4\Lambda^2}+1}.
\label{use}
\end{equation}

\begin{figure}[t]
  \begin{minipage}[t]{0.45\linewidth}
    \centering
    \includegraphics[keepaspectratio, scale=0.6]{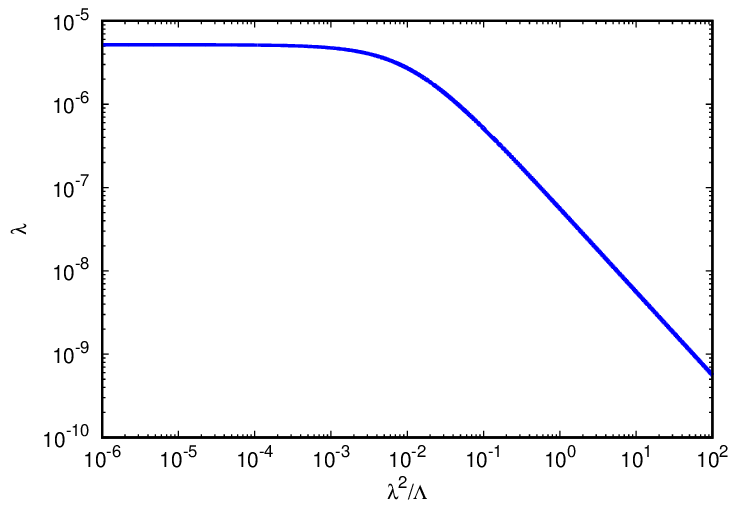}
    \caption{$\lambda$ as a function of $\lambda^2/\Lambda$.}
    \label{fig1}
  \end{minipage}
  \begin{minipage}[t]{0.45\linewidth}
    \centering
    \includegraphics[keepaspectratio, scale=0.6]{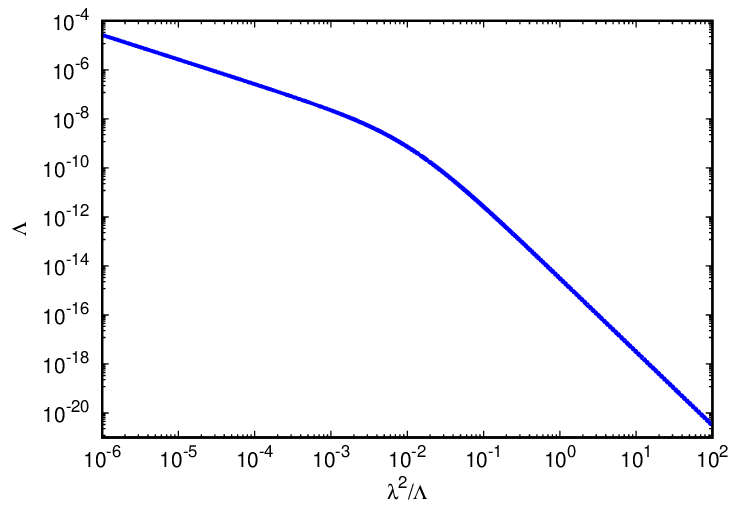}
    \caption{$\Lambda$ as a function of $\lambda^2/\Lambda$.}
  \label{fig2}
  \end{minipage}
\end{figure}

Primordial curvature perturbation can be calculated by using the $\delta N$ formalism \cite{Sasaki:1995aw, Sasaki:1998ug, Lyth:2004gb, Lyth:2005fi, Wands:2000dp}. 
Our $\alpha$ corresponds to $-N$, therefore by using Eqs.~(\ref{a}) and (\ref{alp}) with $\delta \phi=H/2\pi$, we have
\begin{equation}
\zeta=\frac{\partial N}{\partial \phi}\delta \phi =- \frac{\partial \alpha}{\partial \phi}\delta \phi =\frac{\Lambda}{12\pi \lambda}\left[ \cosh^2 \sqrt{\frac{3\lambda^2}{2\Lambda}}\phi-1 \right],
\label{z}
\end{equation}
By Eq.~(\ref{use}) and cosmic microwave background (CMB) normalization $\zeta=5 \times 10^{-5}$, we obtain
\begin{equation}
\frac{\lambda}{12\pi}\left[ \frac{65889\lambda^2}{2\Lambda}+363 \sqrt{\frac{9\lambda^4}{4\Lambda^2}+1}  \right]=5 \times 10^{-5}.
\label{cmb}
\end{equation}
By using $\lambda^2/\Lambda$ as the x-axis, we plot $\lambda$ and $\Lambda$ in Fig.~\ref{fig1} and Fig.~\ref{fig2}.

Another useful identity can be obtained by Eqs.~(\ref{use}) and (\ref{cmb}) (or imposing CMB normalization on Eq.~(\ref{z})),
\begin{equation}
\cosh^2 \sqrt{\frac{3\lambda^2}{2\Lambda}}\phi_{60}-1 = \frac{\lambda}{\Lambda} \times 6\pi \times 10^{-4}.
\label{use2}
\end{equation}

\section{the spectral index and the running spectral index}
\begin{figure}[t]
  \begin{minipage}[t]{0.45\linewidth}
    \centering
    \includegraphics[keepaspectratio, scale=0.9]{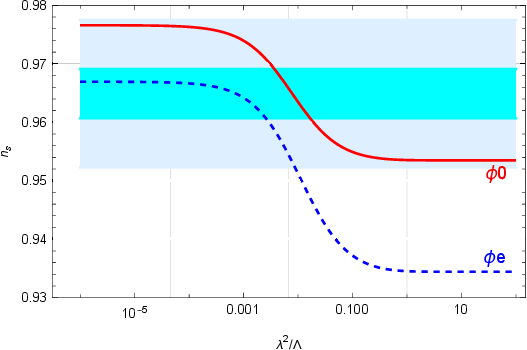}
  \end{minipage}
  \begin{minipage}[t]{0.45\linewidth}
    \centering
    \includegraphics[keepaspectratio, scale=0.9]{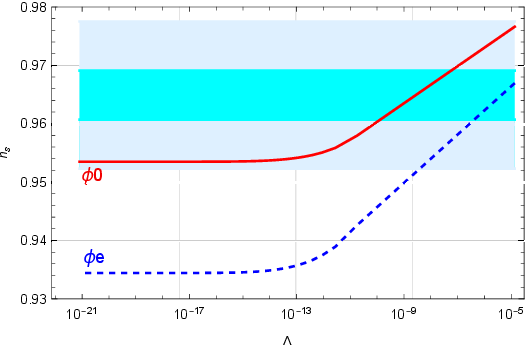}
  \label{fig:nsLmdplot}
  \end{minipage}
    \caption{$n_S$ as a function of $\lambda^2/\Lambda$ (left) and $\Lambda$ (right).
    The blue dashed line corresponds to the case where inflation ends at $\phi_e$, and the red solid line corresponds to the case where inflation ends at $\phi_0$. The latter is discussed in Sec.~\ref{sec:tensortoscalarratio}. Light blue and cyan regions correspond to the allowed parameter spaces of by 68.3 \% (1$\sigma$) and 99.7 \% (3$\sigma$) C.L., respectively \cite{Planck:2018jri}.}
    \label{fig7}
\end{figure}

We can write the spectrum by using Eqs.~(\ref{alp}) and (\ref{z}) as\footnote{More precisely, we should write $P=\langle \zeta^2 \rangle$. However, if we write $\delta \phi=H/2\pi$ instead of $\langle \delta \phi^2 \rangle=(H/2\pi)^2$ for $\zeta=\frac{\partial N}{\partial \phi}\delta \phi$, then $P=\zeta^2$ gives the correct result.}
\begin{equation}
P=\zeta^2 =\frac{\Lambda^2}{144\pi^2 \lambda^2}\left( \frac{9\lambda^4}{\Lambda^2}\alpha^2-1 \right)^2.
\end{equation}
The spectral index is given by
\begin{equation}
n_S \equiv 1+\frac{d\ln P}{d \ln k}=1+\frac{d\ln P}{d\alpha}=1+\frac{\frac{36\lambda^4}{\Lambda^2}\alpha}{\frac{9\lambda^4}{\Lambda^2}\alpha^2-1},
\label{na}
\end{equation}
where $k$ is the comoving wave number of the perturbation.
By using Eqs.~(\ref{alp}), (\ref{co}) and (\ref{use2}), we obtain
\begin{equation}
n_S=1-6366\lambda \left( \frac{363\lambda^2}{2\Lambda}+\sqrt{\frac{9\lambda^4}{4\Lambda^2}+1} \right).
\label{ns}
\end{equation}
This is plotted as the blue dashed lines in Fig.~\ref{fig7}. In the figure, light blue and cyan regions correspond to the allowed parameter spaces of by 68.3 \% (1$\sigma$) and 99.7 \% (3$\sigma$) Confidence Level (C.L.), respectively \cite{Planck:2018jri}.

\begin{figure}[t]
  \begin{minipage}[t]{0.45\linewidth}
    \centering
    \includegraphics[keepaspectratio,scale=0.9]{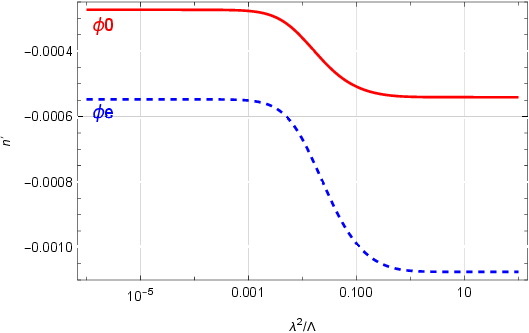}
  \end{minipage}
  \begin{minipage}[t]{0.45\linewidth}
    \centering
    \includegraphics[keepaspectratio,scale=0.9]{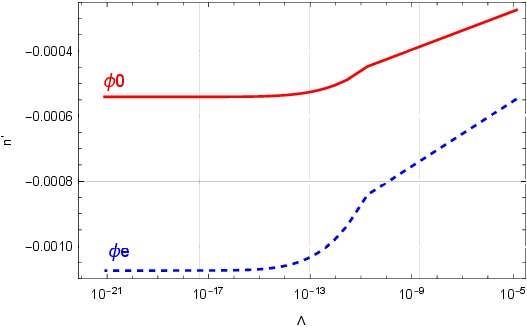}
  \end{minipage}
    \caption{$n^\prime$ as a function of $\lambda^2/\Lambda$ (left) and $\Lambda$ (right). The blue dashed lines correspond to the case where inflation ends at $\phi_e$, and the red solid lines correspond to the case where inflation ends at $\phi_0$. The latter is discussed in Sec.~\ref{sec:tensortoscalarratio}.}
    \label{fig:nplmd2Lmdplot}
\end{figure}

The running spectral index can be obtained by taking one more derivative of $n_S$ given in Eq.~(\ref{na}) with respect to $\alpha$ as
\begin{equation}
n^\prime \equiv \frac{d n_S}{d \ln k} = \frac{dn_S}{d\alpha}=-\frac{\frac{36\lambda^4}{\Lambda^2}\left( \frac{9\lambda^4\alpha^2}{\Lambda^2}+1 \right)}{\left( \frac{9\lambda^4\alpha^2}{\Lambda^2}-1 \right)^2}=-1.91\times 10^4 \frac{\lambda^3}{\Lambda}-2.03\times 10^7\lambda^2,
\end{equation}
where we have used Eqs.~(\ref{alp}) and (\ref{use2}) to obtain the last equality. This is plotted in Fig.~\ref{fig:nplmd2Lmdplot} as the blue dashed lines. 
Observations require $|n^\prime| \lesssim 0.01$ \cite{Planck:2018jri}. Our results of the running spectral index $n^\prime$ are well within the experimental bounds.

\section{the tensor-to-scalar ratio}
\label{sec:tensortoscalarratio}
The tensor-to-scalar ratio $r$ is an experimental parameter that represents primordial gravitational waves generated during inflation. A detection of it may be regarded as a smoking gun of cosmic inflation or a holy grail of experimental search. It is proportional to the ratio of the tensor perturbation and the scalar perturbation.

\begin{figure}[t]
  \begin{minipage}[t]{0.45\linewidth}
    \centering
    \includegraphics[keepaspectratio, scale=0.65]{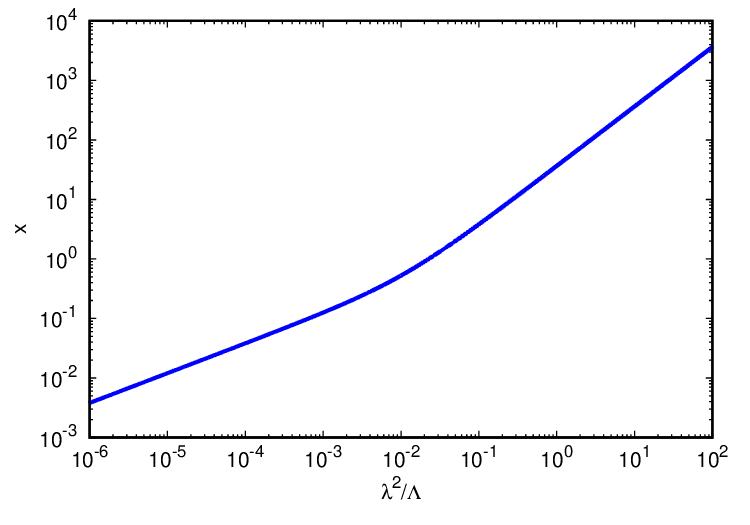}
    \caption{$x$ as a function of $\lambda^2/\Lambda$.}
  \label{fig4}
  \end{minipage}
  \begin{minipage}[t]{0.45\linewidth}
    \centering
    \includegraphics[keepaspectratio, scale=0.65]{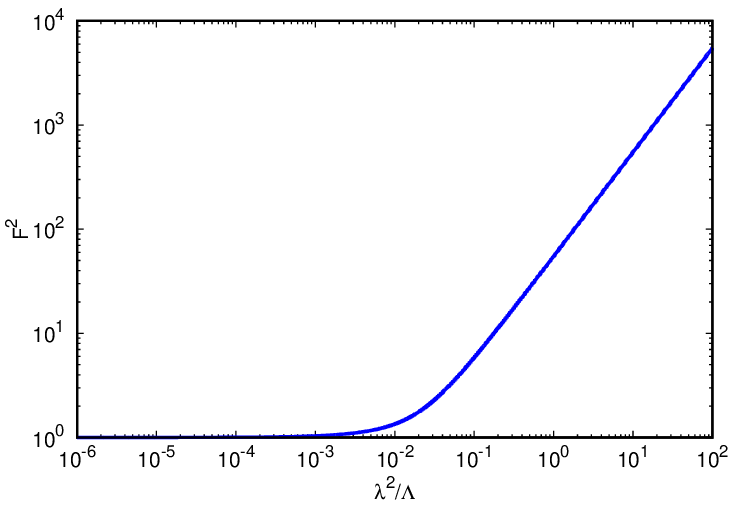}
    \caption{$F^2$ as a function of $\lambda^2/\Lambda$.}
    \label{fig5}
  \end{minipage}
\end{figure}

\begin{figure}[t]
  \begin{minipage}[t]{0.45\linewidth}
    \centering
    \includegraphics[keepaspectratio, scale=0.9]{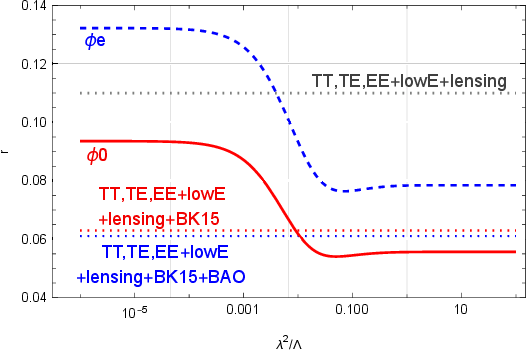}
  \end{minipage}
  \begin{minipage}[t]{0.45\linewidth}
    \centering
    \includegraphics[keepaspectratio, scale=0.9]{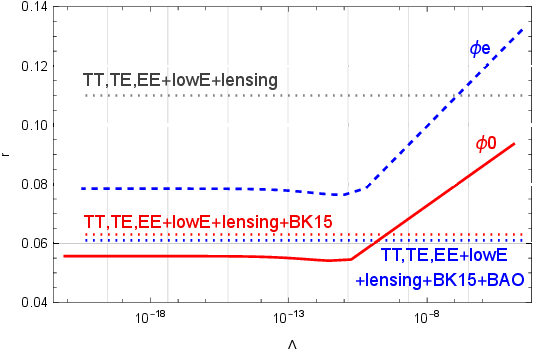}
  \end{minipage}
    \caption{$r$ as a function of $\lambda^2/\Lambda$ (left) and $\Lambda$ (right). The blue dashed lines correspond to the case where inflation ends at $\phi_e$, and the red solid lines correspond to the case where inflation ends at $\phi_0$. The latter is discussed in Sec.~\ref{sec:tensortoscalarratio}. The constraints are from \cite{Planck:2018jri}.}
    \label{fig3}
\end{figure}

\begin{figure}[t]
  \centering
\includegraphics[width=0.6\textwidth]{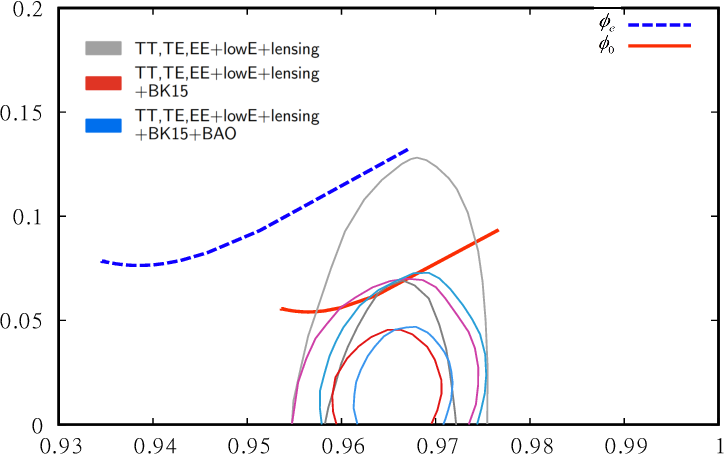}
  \caption{The spectral index and tensor-to-scalar ratio. The constraints are from \cite{Planck:2018jri}. The blue dashed (red) line corresponds to inflation ends at $\phi_e$ ($\phi_0$). The maximum value of predicted $n_S$ recovers the results of \cite{Lin:2023xgs}.}
  \label{fig6}
\end{figure}

The scalar perturbation is represented by
\begin{equation}
A_S^2 = \frac{4}{25}\zeta^2=\frac{H^2\Lambda}{150\pi^2 \lambda^2} \left[ \cosh^2 \sqrt{\frac{3\lambda^2}{2\Lambda}}\phi-1 \right].
\end{equation}
The scalar perturbation generates primordial density perturbations that provide the initial condition for the structure formation of the universe and the CMB temperature anisotropy.
On the other hand, the tensor perturbation in a braneworld is \cite{Langlois:2000ns, Bento:2008yx}
\begin{equation}
A_T^2=\frac{H^2}{50\pi^2}F^2,
\end{equation}
where
\begin{equation}
F^2= \left[ \sqrt{1+x^2}-x^2 \sinh^{-1}\left( \frac{1}{x} \right) \right]^{-1},
\end{equation}
and
\begin{equation}
x \equiv \left( \frac{3H^2}{4\pi \Lambda} \right)^{1/2}.
\end{equation}
From Eqs.~(\ref{a}) and (\ref{use2}), we obtain
\begin{equation}
x^2=7.5\times 10^{-5}\frac{\lambda}{\Lambda}.
\end{equation}
We plot $x$ and $F^2$ in our model as a function of $\lambda^2/\Lambda$ in Fig.~\ref{fig4} and Fig.~\ref{fig5}.
By using Eq.~(\ref{use2}), we obtain
\begin{equation}
\frac{A_T^2}{A_S^2}=\frac{\lambda F^2}{2\pi \times 10^{-4}}.
\end{equation}
The tensor-to-scalar ratio $r$ is defined by
\begin{equation}
r \equiv 16 \frac{A_T^2}{A_S^2}=\frac{8}{\pi}\times 10^4 \lambda F^2.
\end{equation}
We plot $r$ as functions of $\lambda^2/\Lambda$ and $\Lambda$ in Fig.~\ref{fig3} as the blue dashed lines.
We plot the results of $n_S$ and $r$ by 
the blue dashed line to compare with experimental data in Fig.~\ref{fig6}. Unfortunately, the model does not fit the constraints. 

One solution is to assume that inflation ends not at $\phi=\phi_e$, but at $\phi=\phi_0$ with $\phi_0>\phi_e$. For example, let us assume $\Delta \alpha=25$ from $\phi_0$ to $\phi_e$ in the original model, that is 
\begin{equation}
-\frac{\Lambda}{3\lambda^2}\left( \frac{3\lambda^2}{2\Lambda}+\sqrt{\frac{9\lambda^4}{4\Lambda^2}+1} \right)+\frac{\Lambda}{3\lambda^2}\cosh \sqrt{\frac{3\lambda^2}{2\Lambda}}\phi_0=25.
\end{equation}
If inflation ends at $\phi_0$, $\Delta \alpha=25$ is missing. Therefore to have $\Delta \alpha=60$ from $\phi_{60}$ to $\phi_0$, we have to modify the right-hand side of Eq.~(\ref{he}) to $85$.
Using this new value of $\phi_{60}$, we plot the spectral index in Fig.~\ref{fig7}, the tensor-to-scalar ratio in Fig.~\ref{fig3}, and their combination in Fig.~\ref{fig6} as the red solid lines. 
We also plot the running spectral index in Fig.~\ref{fig:nplmd2Lmdplot} as the red solid lines. We can see that experimental constraints can be satisfied by introducing a new parameter $\phi_0$.

\section{conclusion}
In this work, we propose a model of uniform rate inflation on the brane. Due to the complexity of the Hubble parameter given by Eq.~(\ref{brane}) and the corresponding equation of motion, it is pretty remarkable that the equation of motion can be solved analytically without using slow-roll approximation. Furthermore, we carefully calculated the predictions of primordial perturbations, such as the spectral index, the running spectral index, and the tensor-to-scalar ratio. If inflation ends at $\phi_e$, the model could not fit the 
observational constraints. However, if inflation ends at $\phi_0$,  it is possible to fit the 
observational data.

\appendix
\section{stability of the solution}
In this section, we consider a perturbation of $\dot{\phi}$ of the value given by Eq.~(\ref{uni}). Intuitively, we can see from Eqs.~(\ref{eom}) and (\ref{brane}) that if somehow $\dot{\phi}$ is increased (decreased) with the same $V\phi$ and $V\phi^\prime$, $H$ would also increase (decrease). Therefore, the term $3H\dot{\phi}$ in Eq.~(\ref{eom}) is increased (decreased). To compensate for this increasing (decreasing), a negative (positive) $\ddot{\phi}$ appears, and this reduces (enhances) $\dot{\phi}$ towards the value given by Eq.~(\ref{uni}). There is a feedback mechanism. This is not so unexpected since it is known that slow-rolling is an attractor in phase space, and our solution is effectively a slow-roll solution.

Let us discuss it quantitatively. Consider a perturbation $u$ of the solution
\begin{eqnarray}
\dot{\phi} &\rightarrow& \dot{\phi}+u,  \\
\ddot{\phi} &\rightarrow& \ddot{\phi}+\dot{u}  \\
H &\rightarrow& H+\delta H.
\end{eqnarray}
Substituting into Eq.~(\ref{brane}), we have
\begin{equation}
3 \left( H + \delta H \right)^2= \frac{(\dot{\phi}+u)^2}{2}+V(\phi)+\frac{\left[ \frac{(\dot{\phi}+u)^2}{2}+V(\phi) \right]^2}{2\Lambda}.    
\end{equation}
This implies
\begin{equation}
6H\delta H=\dot{\phi}u+\frac{\dot{\phi}u}{\lambda}\left( \frac{\dot{\phi}^2}{2}+V(\phi) \right).    
\end{equation}
By using Eqs.~(\ref{uni}), (\ref{pob}), and (\ref{a}), we obtain
\begin{equation}
\delta H = -\frac{\lambda u}{\sqrt{6\Lambda}}\coth \sqrt{\frac{3\lambda^2}{2\Lambda}}(d-\lambda t).
\end{equation}
Next, take the first-order perturbation of Eq.~(\ref{eom}), we have
\begin{equation}
\dot{u}+\left[\sqrt{\frac{3\Lambda}{2}} \sinh \sqrt{\frac{3\lambda^2}{2\Lambda}}(d-\lambda t)+\lambda^2 \sqrt{\frac{3}{2\Lambda}} \coth \sqrt{\frac{3\lambda^2}{2\Lambda}}(d-\lambda t)\right]u=0.
\end{equation}
The above differential equation can be solved to obtain
\begin{equation}
|u|=b \left| \sinh \sqrt{\frac{3\lambda^2}{2\Lambda}}\phi \right| e^{\frac{\Lambda}{\lambda^2}\cosh \sqrt{\frac{3\lambda^2}{2\Lambda}}\phi},
\end{equation}
where $b$ is an integration constant. Note that $u$ is a small perturbation, and $\phi$ decreases during inflation. Therefore, $u$ drops out very fast. For example, if $\cosh\sqrt{\frac{3\lambda^2}{2\Lambda}}\phi$ drops by a factor of $1/10$. Then $|u|$ drops at least by $0.1 e^{-10}$.

\acknowledgments
This work is supported by the National Science and Technology Council (NSTC) of Taiwan under grant number 112-2112-M-167-001-MY2 (CML). The work is also supported by JSPS Grant-in-Aid for Scientific Research (C) 21K03562
and (C) 21K03583 (KIN). Discussions during the Workshop on Particle and Astroparticle Physics ``Key Interactions Bound for the Invisibles 2023'' were helpful as we started this work.

\end{document}